\documentclass[twocolumn,showpacs,prb,amsmath,amssymb]{revtex4-2}
\usepackage{amsmath,amssymb,color}
\usepackage{bm}
\usepackage{graphicx}
\usepackage{psfrag}
\usepackage{bbold}
\usepackage{bbm}
\newcommand{\bd}{\bm}

\begin{document}

\title{Spontaneous magnon decay in two-dimensional altermagnets \\
}

\author{Niklas Cichutek, Peter Kopietz, and Andreas R\"{u}ckriegel}
  
\affiliation{Institut f\"{u}r Theoretische Physik, Universit\"{a}t
  Frankfurt,  Max-von-Laue Stra{\ss}e 1, 60438 Frankfurt, Germany}

\date{August 14, 2025}

 \begin{abstract}

We show that magnons in two-dimensional altermagnets can spontaneously decay at zero  temperature. The decay rate is determined by quantum fluctuations and scattering processes involving the decay of a single magnon into three. These processes are kinematically allowed due to the
convexity of the altermagnetic magnon dispersion.
For small wavevectors $k$ the decay rate is proportional to $k^5$ with a direction-dependent prefactor which is maximal along the diagonals of the Brillouin zone. Moreover, 
for a given momentum only magnons with one specific chirality can spontaneously decay.

\end{abstract}


\maketitle

\section{Introduction}

The spontaneous decay of quasiparticles at zero temperature is a fundamental manifestation of quantum mechanics  in condensed matter.
In general, this is only possible if 
the ground state exhibits zero point (quantum) fluctuations 
and the quasiparticle  dispersion $\omega_{\bd{k}}$ is a 
convex function of the momentum $\bd{k}$ so that the decay of a single quasiparticle into two or more quasiparticles is kinematically allowed. A well-known example where both conditions are satisfied is the quasiparticle decay  in the
superfluid ground state of interacting bosons \cite{Lifshitz80}. In this case the low-energy quasiparticles are phonons with long-wavelength dispersion
 $\omega_{\bd{k}} = c k + \alpha k^3$ 
where the sound  velocity
$c$ and the coefficient $\alpha$ of the cubic non-linearity
are positive, implying that  $\omega_{\bd{k}}$ is a convex function of
$k = | \bd{k} |$.
The dominant scattering process which determines the 
so-called Beliaev damping \cite{Beliaev58} of the phonons at zero temperature involves the
decay of one phonon into two.
In three dimensions the corresponding decay rate is proportional to
$k^5$ at long wavelengths \cite{Lifshitz80},  which generalizes to
$k^{2d-1}$ in $d > 1$ dimensions \cite{Sinner09}.  

In the above example the convexity of the quasiparticle dispersion 
guarantees that the relevant decay process satisfies energy and momentum conservation.
Conversely, quasiparticles with a  concave dispersion cannot spontaneously decay to zero temperature,
even if the ground state exhibits quantum fluctuations.
This is the reason why magnons in 
quantum Heisenberg antiferromagnets with nearest-neighbor 
exchange on a square lattice are stable
at zero temperature \cite{Harris71,Zhitomirsky13}.
In this work we show that in some sectors of the Brillouin zone 
such a kinematic constraint does not apply to magnons
in the recently discovered altermagnets \cite{Smejkal22a,Smejkal22b,Smejkal23}, so that 
magnons in altermagnets can in fact spontaneously decay. For simplicity we focus here on two-dimensional altermagnets and show that the corresponding decay rate is dominated by the decay of one magnon into three.  For small momenta the  decay rate is proportional to $k^5$ and depends on the direction in momentum space \cite{footnote_Eto}.
Interestingly, magnon decay in altermagnets is found to be {\it{chirality selective}}
in the sense that for fixed $\bd{k}$ only the magnons 
with one specific chirality 
can spontaneously decay while the magnons with the opposite chirality 
are stable.
Spontaneous magnon decay in altermagnets has recently been discussed in Ref.~[\onlinecite{GarciaGaitan24}]; however, an analytic understanding of the underlying physical mechanism has not been obtained.

\section{Minimal spin model}

\begin{figure}[tb]
 \begin{center}
  \centering
 \includegraphics[width=0.3\textwidth]{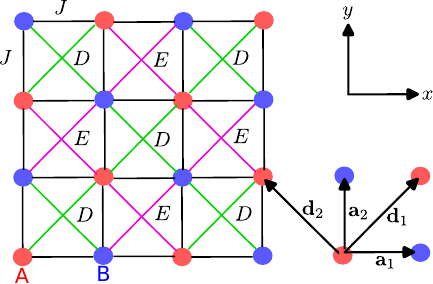}
   \end{center}
  \vspace{-4mm}
  \caption{%
Illustration of the spin model for an altermagnet on a square lattice defined in Eq.~(\ref{eq:modelc}). The lattice is divided into two sublattices $A$ and $B$. The exchange interaction $J$ connects all pairs of nearest-neighbor spins, while the two types of couplings $D$ and $E$ connect the spins accross the diagonals of plaquettes in a checkerboard pattern.
}
\label{fig:models3}
\end{figure}

Starting point of our investigation is a minimal spin model
for two-dimensional altermagnets on a square lattice with Hamiltonian \cite{Brekke23}
\begin{align}
 {\cal{H}}  & = J \sum_{\bd{R} }
  \bd{S}_{\bd{R}} \cdot ( \bd{S}_{ \bd{R} + \bd{a}_1}  +  \bd{S}_{ \bd{R} + \bd{a}_2} )
  \nonumber
  \\
  & 
   + 
   \sum_{\bd{R} \in A} \bd{S}_{\bd{R}} \cdot  (
     D  \bd{S}_{\bd{R} + \bd{d}_1 }
     +
    E  \bd{S}_{\bd{R} + \bd{d}_2 })
   \nonumber
   \\
   &
   +  
   \sum_{\bd{R} \in B}  \bd{S}_{\bd{R}} \cdot (
     E  \bd{S}_{\bd{R} + \bd{d}_1 }
     + D  \bd{S}_{\bd{R} + \bd{d}_2 } ).
  \label{eq:modelc}
 \end{align}
where $\bd{S}_{\bd{R}}$ are spin-$S$ operators localized at sites
$\bd{R}$  of a square lattice with $N$ sites, lattice spacing $a$ and basis vectors
$\bd{a}_1 = a \hat{\bd{x}}$ 
and $\bd{a}_2 = a \hat{\bd{y}}$.
We have divided the lattice into two sublattices $A$ and $B$ as shown in Fig.~\ref{fig:models3}.
The antiferromagnetic exchange coupling  $J > 0$ connects all pairs of nearest-neighbor spins
while  the diagonal coupling $D$ connects spins across the diagonals 
 $\bd{d}_1 =
a ( \hat{\bd{x}} + \hat{\bd{y}} )$ and  $\bd{d}_2 =
a ( - \hat{\bd{x}} + \hat{\bd{y}} )$ on 
half of the plaquettes. On the diagonals of the other plaquettes
the  spins are connected by a different coupling
$E$ such that the distinct plaquettes form a checkerboard pattern as illustrated in 
Fig.~\ref{fig:models3}. 
Physically, $D$ and $E$ describe next-nearest neighbor exchange 
mediated by two different types of non-magnetic atoms located at the centers of the corresponding plaquettes.
By setting $E=0$ we obtain the model proposed in Ref.~[\onlinecite{Consoli24}].

\begin{figure}[htb]
 \begin{center}
  \centering
\vspace{7mm}
 \includegraphics[width=0.4\textwidth]{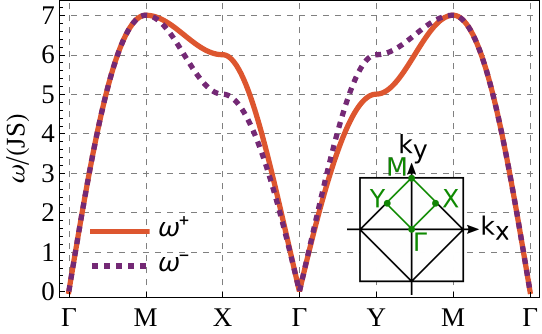}
   \end{center}
  \caption{%
Magnon dispersions $\omega_{\bd{k}}^{\pm} $ defined in Eq.~(\ref{eq:magdisp}) for the path in the Brillouin zone
shown in the inset where the M-point represents the wavevector
$( 0, \pi/a)$. 
The graph is for $D/J = - 1/2$ and $E/J = -1/4$. 
}
\label{fig:dispersions}
\end{figure}

Assuming that the next-nearest neighbor couplings $D$ and $E$ are small compared with $J$,
the spin configuration in the classsical ground state is the usual
Ne\'{e}l state where spins on the two sublattices are collinear and point in opposite directions.
The magnon spectrum can then be obtained by  means of the usual steps:
after bosonizing the spin operators via the Holstein-Primakoff transformation \cite{Holstein40}, 
we transform to momentum space and then use a Bogoliubov transformation to completely diagonalize the Hamiltonian. The quadratic part of the resulting magnon Hamiltonian is
\begin{equation}
 {\cal{H}}_2 =  \sum_{\bd{k}} \left( 
 \omega_{\bd{k}}^{+} \alpha^{\dagger}_{\bd{k}}   \alpha_{\bd{k}}
+ 
 \omega_{\bd{k}}^{-} \beta^{\dagger}_{\bd{k}}   \beta_{\bd{k}}
 \right),
 \end{equation}
where the sum is over the reduced Brillouin zone associated with the $A$-sublattice
containing $N/2$ sites, $\alpha_{\bd{k}} $ and $\beta_{\bd{k}}$ are canonical annihilation operators for magnons with chirality $p = \pm$ and dispersions 
 \begin{align}
 \omega_{\bd{k}}^{p} & = 4 J S \epsilon_{\bd{k}} 
 + p   ( E - D ) 2 S \sin ( k_x a ) \sin ( k_y a ),
  \label{eq:magdisp}
 \end{align}
with
 \begin{subequations}
 \begin{align}
 \epsilon_{\bd{k}} & = \sqrt{ ( 1 - \eta_{\bd{k}} )^2 - \gamma_{\bd{k}}^2 },
 \\
 \eta_{\bd{k}} & =   \frac{D + E}{2J}  [ 1 - \cos ( k_x a ) \cos ( k_y a ) ],
 \\
 \gamma_{\bd{k}} & = \frac{1}{2} \left[ \cos ( k_x a ) + \cos ( k_y a ) \right].
 \end{align} 
 \end{subequations}
A graph of the magnon dispersions is shown in
Fig.~\ref{fig:dispersions}.
If $k$ is small compared with the inverse lattice spacing the magnon
dispersions can be approximated by
 \begin{align}
 \omega^{\pm}_{\bd{k}} & = c | \bd{k} | \pm \frac{ k_x k_y}{m} + {\cal{O}} ( k^3 ),
 \label{eq:omegaexp}
 \end{align}
with magnon velocity $c$ and inverse mass $1/m$ given by
 \begin{align}
 c & = 
 2 \sqrt{2} J S a \sqrt{ 1 - \frac{ D + E}{J} }, \; \; \; \; \; 
   \frac{1}{m}  =
  2 ( E - D ) S  a^2 .
 \end{align}

\section{Spontaneous magnon decay}

\begin{figure}[htb]
 \begin{center}
  \centering
 \includegraphics[width=0.48\textwidth]{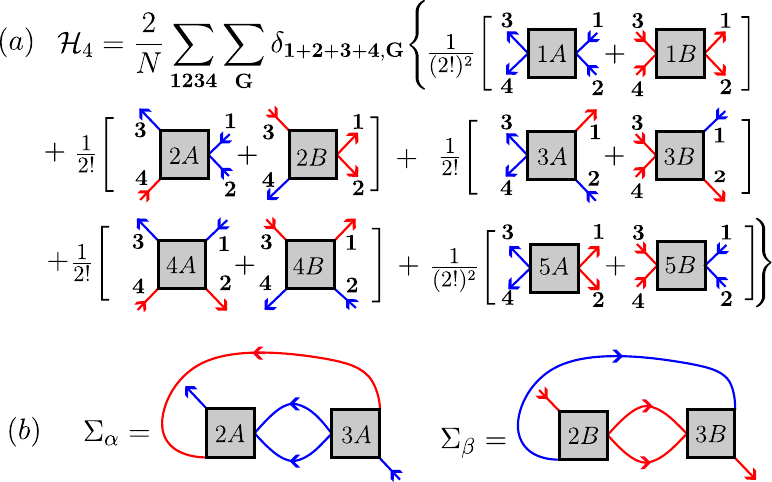}
   \end{center}
  \caption{%
(a): Graphical  representation of the bosonized two-body Hamiltonian ${\cal{H}}_4$ of an altermagnet in terms of magnon operators.
Here bold numbers $\bd{1}, \bd{2}, \bd{3} , \bd{4}$ represent the 
lattice momenta $\bd{k}_{1}$, $\bd{k}_2$, $\bd{k}_3$, $\bd{k}_4$ which
are conserved up to a vector $\bd{G}$ of the reciprocal lattice associated with the $A$-sublattice, i.e., $e^{i \bd{G} \cdot \bd{R} } =1$ for
$\bd{R} \in A$. 
Outgoing/incoming  blue arrows represent $\alpha^{\dagger}_{- \bd{k} }$/$\alpha_{\bd{k}}$,
while outgoing/incoming red arrows represent $\beta^{\dagger}_{- \bd{k}}$/$\beta_{\bd{k}}$.
Arrows entering or leaving a vertex at the same point indicate that the vertex is symmetric with respect to the exchange of the corresponding momenta.
 (b): Diagrammatic representation of the two-loop corrections to the magnon self-energies at zero temperature. Internal lines with arrows represent imaginary-frequency magnon propagators.
}
\label{fig:h4}
\end{figure}

For the calculation of the 
decay rate of the magnons we have to express the quartic part
${\cal{H}}_4$  of the bosonized spin Hamiltonian in terms of magnon operators $\alpha_{\bd{k}}$ and $\beta_{\bd{k}}$ and the
corresponding creation operators. A graphical representation of the resulting
two-body Hamiltonian is shown in Fig.~\ref{fig:h4} (a).
The various scattering processes can be parametrized in terms of
$10$ distinct interaction vertices $\Gamma^{(1A)}, \Gamma^{(1B)}, \ldots, \Gamma^{(5A)}, \Gamma^{(5B)}$ 
represented by the shaded squares. 
To exhibit the symmetries of the vertices we 
normal-order ${\cal{H}}_4$ with respect to the $\alpha$-operators and 
anti-normal order with respect to the 
$\beta$-operators \cite{Kopietz90}. It is also convenient to introduce proper combinatorial factors to take into account the
symmetries of the vertices with  respect to permutations of external legs of the same type.
It turns out that the spontaneous magnon decay rates 
are determined by
scattering processes where one magnon decays into three ($1 \rightarrow 3$ process) or three magnons merge into one ($3 \rightarrow 1$ process).
These processes are determined by the vertices
$\Gamma^{(2A)}, \Gamma^{(2B)}, \Gamma^{(3A)}$ and $\Gamma^{(3B)}$ 
in the second line of Fig.~\ref{fig:h4}~(a)
involving either one creation- and three annihilation operators, or one annihilation- and three creation operators.
For the calculation of the  
spontaneous decay rate of the $\alpha$-magnons 
we only need 
the vertex $\Gamma^{(2A)}$ which for general  Umklapp momentum 
$\bd{G}$ is explicitly given by 
 \begin{widetext}
 \begin{align}
  \Gamma^{(2A)}_{\bd{G}} ( \bd{3} ; \bd{4}; \bd{2} , \bd{1} )   =
 2 J  & \Bigl[  \bigl( u_{\bd{3}} u_{\bd{4}} \gamma_{ \bd{2} + \bd{4} } + v_{\bd{3}} v_{\bd{4}} \gamma_{\bd{2} + \bd{3}} \bigr) v_{\bd{2}} u_{\bd{1}}
 + \bigl( u_{\bd{3}} u_{\bd{4}} \gamma_{ \bd{1} + \bd{4} } + v_{\bd{3}} v_{\bd{4}} \gamma_{\bd{1} + \bd{3}} \bigr)  u_{\bd{2}} v_{\bd{1}}
 \nonumber
 \\
 &  
 + s_{\bd{G}} \bigl( v_{\bd{3}} v_{\bd{4}} \gamma_{ \bd{2} + \bd{4} } + u_{\bd{3}} u_{\bd{4}} \gamma_{\bd{2} + \bd{3}} \bigr) u_{\bd{2}} v_{\bd{1}}
 + s_{\bd{G}} \bigl( v_{\bd{3}} v_{\bd{4}} \gamma_{ \bd{1} + \bd{4} } + u_{\bd{3}} u_{\bd{4}} \gamma_{\bd{1} + \bd{3}} \bigr)  v_{\bd{2}} u_{\bd{1}}
 \nonumber
 \\
 & 
 - u_{\bd{3}} v_{\bd{4}} \bigl(  u_{\bd{2}} v_{\bd{1}}  \gamma_{\bd{1}}
  +  v_{\bd{2}} u_{\bd{1}}  \gamma_{\bd{2}}   \bigr)
  - \bigl( v_{\bd{3}} v_{\bd{4}} \gamma_{\bd{3}} + u_{\bd{3}} u_{\bd{4}} \gamma_{\bd{4}}
   \bigr) u_{\bd{2}} u_{\bd{1}}
  -  s_{\bd{G}} v_{\bd{3}} u_{\bd{4}} \bigl(  v_{\bd{2}} u_{\bd{1}}  \gamma_{\bd{1}}
  +  u_{\bd{2}} v_{\bd{1}}  \gamma_{\bd{2}}  \bigr)
   \nonumber
   \\
   &
    - s_{\bd{G}} \bigl( u_{\bd{3}} u_{\bd{4}} \gamma_{\bd{3}} + v_{\bd{3}} v_{\bd{4}} \gamma_{\bd{4}}
   \bigr) v_{\bd{2}} v_{\bd{1}}
   \Bigr]
   - u_{\bd{3}} v_{\bd{4}} u_{\bd{2}} u_{\bd{1}} W^A_{ \bd{3} \bd{4} ; \bd{2} \bd{1} }
  - s_{\bd{G}} v_{\bd{3}} u_{\bd{4}} v_{\bd{2}} v_{\bd{1}} W^B_{ \bd{3} \bd{4} ; \bd{2} \bd{1} }.
  \label{eq:Gamma2}
 \end{align}
\end{widetext}
Here the usual Bogoliubov coefficients are given by
 \begin{equation}
 u_{\bd{k}}  = \sqrt{ \frac{1 - \eta_{\bd{k}} +
  \epsilon_{\bd{k}} }{2 \epsilon_{\bd{k}} }},
  \; \; \;
  v_{\bd{k}} = s_{\bd{k}}  \sqrt{ \frac{1 - \eta_{\bd{k}} -
  \epsilon_{\bd{k}} }{2 \epsilon_{\bd{k}} }},
  \end{equation}
with $s_{\bd{k}} = {\rm sign} \gamma_{\bd{k}}$.
The last two terms in Eq.~(\ref{eq:Gamma2}) 
depend on the 
altermagnetic couplings $D$ and $E$ as follows,
 \begin{align}
 W^X_{\bd{3} \bd{4}; \bd{2} \bd{1}} & =
 V^X_{\bd{1} + \bd{3}} + V^X_{\bd{1} + \bd{4}} +
 V^X_{\bd{2} + \bd{3}} + V^X_{\bd{2} + \bd{4}} 
 \nonumber
 \\
 & -V^X_{\bd{1} } - V^X_{\bd{2}} -
 V^X_{\bd{3}} - V^X_{ \bd{4}} , \; \; \; \;  \; \;  X = A,B,
 \end{align}
where
 $V^A_{\bd{k}} = D \cos ( \bd{k} \cdot \bd{d}_1 ) + 
E \cos ( \bd{k} \cdot \bd{d}_2 )$ and
$V^B_{\bd{k}} = E \cos ( \bd{k} \cdot \bd{d}_1 ) + 
D \cos ( \bd{k} \cdot \bd{d}_2 )$.
Note that the hermiticity of the Hamiltonian implies that the
vertex $\Gamma^{(3A)}$ can be obtained from
$\Gamma^{(2A)}$ via complex conjugation,
 \begin{equation}
 \Gamma^{(3A)}_{\bd{G}} ( \bd{3} , \bd{4} ; \bd{2} ;   \bd{1} )
 = [ \Gamma^{(2A)}_{\bd{G}} ( - \bd{2} ; - \bd{1} ; -\bd{3} ,  -\bd{4} )
  ]^{\ast}.
   \label{eq:hermiticity}
 \end{equation}

\begin{figure}[htb]
 \begin{center}
  \centering
 \includegraphics[width=0.45\textwidth]{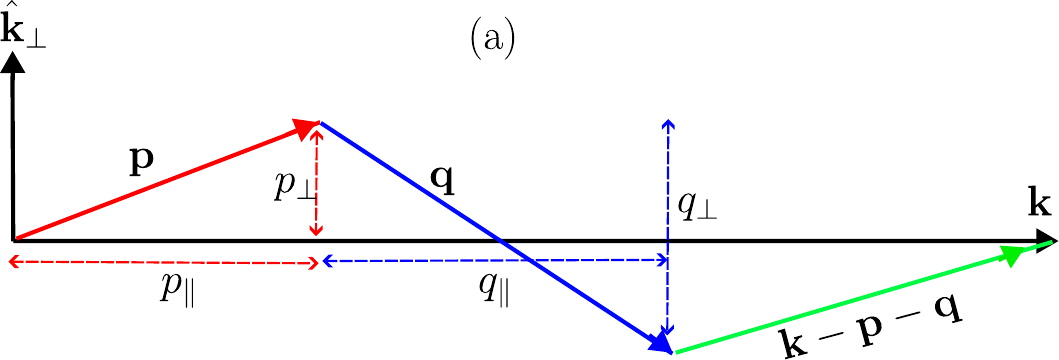}
 \vspace{7mm}
\\
 \includegraphics[width=0.47\textwidth]{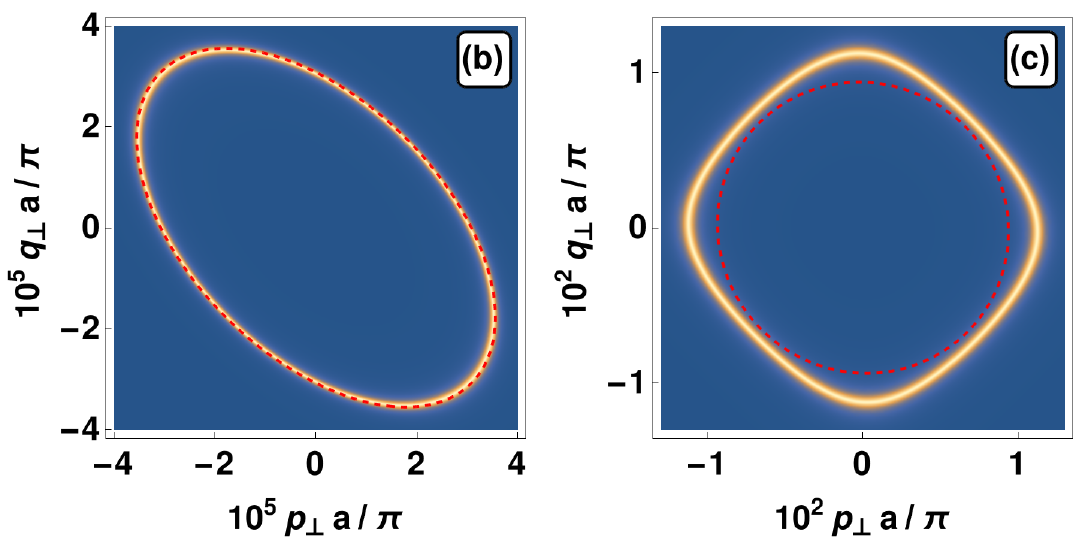} 
   \end{center}
  \caption{%
(a) For small non-linearity energy and momentum conservation imply that the decay of one magnon with momentum $\bd{k}$ into three magnons
with momenta $\bd{p}$,  $\bd{q}$,  and $\bd{k} - \bd{p} - \bd{q}$ is only possible if $\bd{p}$ and $\bd{q}$ are almost aligned with $\bd{k}$.
(b) Elliptic scattering surface in the  $p_{\bot}$-$q_{\bot}$ plane associated with spontaneous magnon decay. The red dashed line
represents the solution of Eq.~(\ref{eq:ellipse}) which is valid 
for small $ka$ and to leading order in
$p_{\bot} /k $ and $q_{\bot} /k$.  The yellow curve  is obtained from the numerical solution of the original Eq.~(\ref{eq:scatsurf}).
We have chosen the following parameters: $D/J = -1/2$, $E/J = -1/4$, $k_x a = k_y a = 0.003 \pi /\sqrt{2}$, $p_{\parallel} a = q_{\parallel} a = 0.001 \pi$. (c) For larger wavevectors
the shape of the scattering surface deviates from an ellipse.
}
\label{fig:forwardscat}
\end{figure}

Given the magnon Hamiltonian ${\cal{H}}_2 + {\cal{H}}_4$ we may now calculate the magnon self-energies using the imaginary time formulation of diagrammatic perturbation theory. 
 To obtain the leading contribution to the
 magnon damping, we have to calculate all diagrams up to second order
 in the interaction ${\cal{H}}_4$. Let us focus on the
 diagonal self-energy $\Sigma_{\alpha} ( \bd{k} , i \omega )$ of the
 $\alpha$-magnon with chirality $ p = + $, where $i \omega$ is a bosonic Matsubara frequency.
 Due to the rather complicated interaction
 ${\cal{H}}_4$ eight distinct second-order diagrams
 contribute to $\Sigma_{\alpha} ( \bd{k} , i \omega )$ which depend on various combinations of the interaction vertices. Fortunately, in the
 limit of vanishing temperature $T \rightarrow 0$ only the first diagram shown
in Fig.~\ref{fig:h4} (b) survives. 
This diagram has the unique property that the arrows attached to all external and internal
lines point into the same direction.
After evaluating the relevant frequency sums at finite $T$ we find that the contribution of this  diagram  to the 
 self-energy of the $\alpha$-magnon 
can be written as
 \begin{align}
 & \Sigma_{\alpha} ( \bd{k} , i \omega )  =  \frac{1}{2} \left( \frac{2}{N} \right)^2
  \sum_{ \bd{p}, \bd{q}} 
  \left[ 1 - e^{ - (  \omega^+_{\bd{p}} + \omega^+_{\bf{q}}     
  + \omega^-_{\bd{k} - \bd{p} - \bd{q}} ) / T} \right]  
 \nonumber
 \\
 & \times   W( \bd{p} , \bd{q} ; \bd{k} )  \frac{   (1 + n_{\bd{p}}^+ ) ( 1 + n_{\bd{q}}^+ )
 (1 + n^-_{\bd{k} - \bd{p} - \bd{q} } ) }{ i \omega
  - \omega^+_{\bd{p}} - \omega^+_{\bd{q}} - \omega^-_{\bd{k} - \bd{p} - \bd{q} } } ,
  \label{eq:selfzeroplus} 
  \end{align}
where
 $ n^{\pm}_{\bd{p}}  =1 / [e^{ \omega^{\pm}_{\bd{p}} / T } -1 ]$.
The symmetry (\ref{eq:hermiticity}) implies that the relevant product of  vertices is positive,
 \begin{align}
   W ( \bd{p} , \bd{q} ; \bd{k} ) & =
{\Gamma}^{(2A)}_0 ( - \bd{k};  \bd{k} - \bd{p} - \bd{q}  ; \bd{p} ,  \bd{q} )
 \nonumber  
\\
 & \times
   {\Gamma}^{(3A)}_0 (   - \bd{p} , - \bd{q} ;   \bd{k};  \bd{p} + \bd{q} - \bd{k}  )
 \nonumber
 \\
 & =     | {\Gamma}^{(2A)}_0 ( - \bd{k};  \bd{k} - \bd{p} - \bd{q}  ; \bd{p} ,  \bd{q} ) |^2,
  \label{eq:Wmat}
    \end{align}
where we have assumed that the external momentum $\bd{k}$ 
is sufficiently small 
so that we may set $\bd{G} =0$ in all vertices. For $T \rightarrow 0$
Eq.~(\ref{eq:selfzeroplus}) yields for the damping of the $\alpha$-magnons on resonance
 \begin{align}
 & \gamma_{\alpha} ( \bd{k} )  = - {\rm Im} \Sigma_{\alpha} ( {\bd{k}}, \omega_{\bd{k}}^+ + i 0^+ ) =
 \nonumber
 \\
 & 
 \frac{\pi}{2} \left( \frac{2}{N} \right)^2
  \sum_{ \bd{p}, \bd{q}}  
  W ( \bd{p} , \bd{q} ; \bd{k} )
  \delta ( \omega_{\bd{k}}^{+} - \omega_{\bd{p}}^{+} 
   -  \omega^{+}_{\bd{q}} -
   \omega^{-}_{  \bd{k} - \bd{p} -  \bd{q}} ).
   \label{eq:damp1}
 \end{align}
Spontaneous magnon decay is kinematically allowed if 
the equation for the scattering surface,
\begin{equation}
\omega_{\bd{k}}^{+} = \omega_{\bd{p}}^{+} 
    + \omega^{+}_{\bd{q}} +
   \omega^{-}_{  \bd{k} - \bd{p} -  \bd{q}},
   \label{eq:scatsurf}
 \end{equation}
has non-trivial solutions.
A similar problem of finding the scattering surface
for the zero-temperature decay of quasiparticles with linear dispersion plus 
 cubic non-linearity has recently been discussed by O'Brien and Sushkov \cite{OBrien20}.
A crucial observation  is that for small non-linearity the loop momenta
$\bd{p}$ and $ \bd{q}$ are almost aligned with the direction of the external momentum $\bd{k}$ as shown in Fig.~\ref{fig:forwardscat} (a).
It is then convenient to project $\bd{p} = p_{\parallel} \hat{\bd{k}} + {p}_{\bot} \hat{\bd{k}}_{\bot}$ and $\bd{q} = q_{\parallel} \hat{\bd{k}} + {q}_{\bot} \hat{\bd{k}}_{\bot}  $ onto the direction
$\hat{\bd{k}} = \bd{k} / | \bd{k} |$ of the external momentum,  
where $\hat{\bd{k}}_{\bot}$ is a unit vector perpendicular to ${\bd{k}}$.
Assuming that all momenta  are small compared with the inverse lattice spacing,
we may approximate the magnon dispersions in Eq.~(\ref{eq:scatsurf})
by the small-momentum expansion
(\ref{eq:omegaexp}).
Anticipating that
the transverse components $p_{\bot}$ and $q_{\bot}$ are parametrically smaller than the longitudinal components
$p_{\parallel}$, $ q_{\parallel}$ and $k - p_{\parallel } - q_{\parallel}$, we may further expand up to quadratic order in
$p_{\bot}$ and $q_{\bot}$. 
For $k \lesssim   {\rm min } \{ 1/a, |mc| \}$ the equation (\ref{eq:scatsurf}) 
for the scattering surface then reduces to
\begin{align}
 &  \frac{ p_{\bot}^2}{p_{\parallel}} + \frac{q_{\bot}^2}{ q_{\parallel}}
  + \frac{ ( p_{\bot} + q_{\bot} )^2 }{ k - p_{\parallel} - q_{\parallel} } 
 = \frac{4 \hat{k}_x \hat{k}_y   }{mc}   ( k - p_{\parallel} ) ( k - q_{\parallel} ).
  \label{eq:ellipse}
 \end{align}
Keeping in mind that momentum conservation requires
$0 < p_{\parallel} < k$, $0< q_{\parallel} < k $, and $0 < k - p_{\parallel} - q_{\parallel} < k $, the quadratic form on the
left-hand side of Eq.~(\ref{eq:ellipse}) is positive definite.
For   $m > 0$ this implies that  Eq.~(\ref{eq:ellipse}) has only a 
solution
if $ \hat{k}_x \hat{k}_y > 0$; i.e., $ \bd{k}$ must lie  in the first or in the  third quadrant 
of the Brillouin zone. For fixed $p_{\parallel}$ and $q_{\parallel}$
the solution of Eq.~(\ref{eq:ellipse}) is then an ellipse
in the $p_{\bot}$-$q_{\bot}$ plane, as illustrated in 
Fig.~\ref{fig:forwardscat} (b). The elliptic approximation to the scattering surface is accurate to leading order in $ka$ and $k /|mc|$. For larger wavevectors the shape of the scattering surfaces deviates from an ellipse, as illustrated in Fig.~\ref{fig:forwardscat} (c).

Let us now calculate the leading momentum-dependence of the decay rate for $k \rightarrow 0$. In this long-wavelength limit we may approximate the vertex 
in Eq.~(\ref{eq:Wmat}) by
 \begin{align}
&  {\Gamma}^{(2A)}_0 ( \bd{3} ; \bd{4}; \bd{2} , \bd{1} ) =
\nonumber 
\\
  & J 
   \left[  \sqrt{ \frac{k_3 k_4}{k_2 k_1}}
    ( 1 + \hat{\bd{k}}_3 \cdot \hat{\bd{k}}_4 )
    -  \sqrt{ \frac{k_2 k_1}{k_3 k_4}}
    ( 1 - \hat{\bd{k}}_2 \cdot \hat{\bd{k}}_1 )
     \right].
 \end{align}
Note that in this limit the vertex is independent of the altermagnetic couplings $D$ and $E$, 
so that magnon decay in altermagnets is due to the specific kinematics of the band splitting.
Keeping in mind that for the scattering geometry of interest
the loop momenta
$\bd{p}$ and $\bd{q}$ are almost parallel to the external momentum
$\bd{k}$, the relevant squared matrix element can be approximated by
 \begin{align}
  W ( \bd{p} , \bd{q} ; \bd{k} )  & = \frac{J^2}{4} \Biggl[
\sqrt{ \frac{ p_{\parallel}q_{\parallel} }{k ( k- {p}_{\parallel} - {q}_{\parallel} ) }}
 \left( \frac{ p_{\bot}}{p_{\parallel} } - \frac{q_{\bot}}{q_{\parallel} } \right)^2
 \nonumber
 \\
 & -    \sqrt{ \frac{ k ( k- {p}_{\parallel} - {q}_{\parallel} ) }{ p_{\parallel} q_{\parallel}}}
 \frac{ ( p_{\bot} + q_{\bot} )^2 }{  ( k - p_{\parallel} - q_{\parallel} )^2 }
 \Biggr]^2.
  \label{eq:Wlong}
\end{align}
Substituting this into Eq.~(\ref{eq:damp1})
and using the elliptic approximation (\ref{eq:ellipse})
for the scattering surface, we may scale out the momentum dependence 
by setting 
 $p_{\parallel}  = k x_1$, $q_{\parallel}  = k x_2 $, and 
 $p_{\bot} =  \sqrt{2 k^3 / |mc|}  y_1$,
 $q_{\bot}  = \sqrt{  2 k^3 / | mc |}  y_2$.
Note that the transverse momenta scale as $k^{3/2}$, whereas the
longitudinal momenta are of order $k$ so that for sufficiently small $k$ our assumption that
$p_{\bot}$ and $q_{\bot}$ are small compared with
$p_{\parallel}$ and $q_{\parallel}$ is justified.
The transverse integration over
$y_1$ and $y_2$ can then be carried out analytically, while the
remaining two-dimensional integral over $x_1$ and $x_2$ can be evaluated numerically. Our final result for the decay rate of
$\alpha$-magnons is \cite{footnote_real_part}
\begin{align}
 \gamma_{\alpha} ( {\bd{k}}) & = 0.0135 \;  \Theta \Bigl( \frac{ k_x k_y}{m} \Bigr)  \frac{ J^2 a^4 | \bd{k} | k_x^2 k_y^2 }{m^2 c^3} 
  + {\cal{O}} ( k^6 ),
  \label{eq:dampres}
 \end{align}
where $\Theta (x) = 1$ for $ x > 0$ and $\Theta (x ) = 0$ for $x < 0$. 
The decay rate of the $\beta$-magnons with chirality $ p = - $ can be obtained by substituting
$\Theta ( k_x k_y /m) \rightarrow \Theta (- k_x k_y /m )$ in Eq.~(\ref{eq:dampres}).

\section{Conclusions}

\begin{figure}[htb]
 \begin{center}
  \centering
\includegraphics[width=0.25\textwidth]{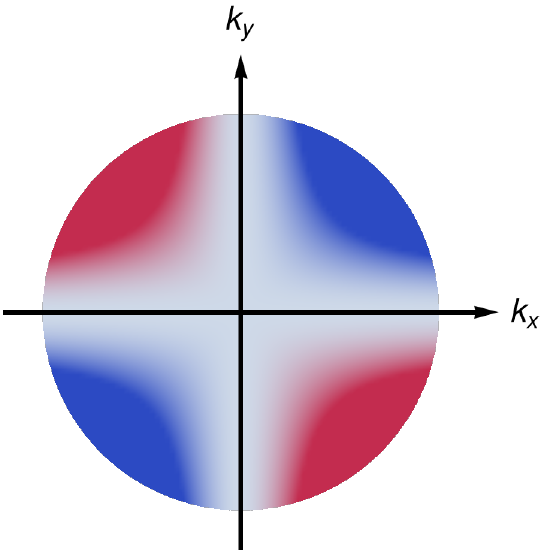}
   \end{center}
  \caption{%
Illustration of the chirality selectivity of the spontaneous magnon decay rate
in two-dimensional altermagnets for small momenta.
The intensity of the blue shading is proportional to
the spontaneous decay rate $\gamma_{\alpha}
(\bd{k} )$ of the $\alpha$-magnon given in Eq.~(\ref{eq:dampres}), while the
red shading represents the corresponding  decay rate $\gamma_{\beta} ( \bd{k} )$ of the $\beta$-magnon. We only show the long-wavelength regime  $ | \bd{k}  | \ll 1/a $ where the approximations leading to 
Eq.~(\ref{eq:dampres}) are expected to be  accurate. 
}
\label{fig:dampres}
\end{figure}

We conclude that  spontaneous magnon decay in two-dimensional altermagnets is chirality selective:
for positive $m$ the $\alpha$-magnons can spontaneously decay only in the
first and third quadrant of the Brillouin zone, while the $\beta$-magnons can decay only 
in the second and fourth quadrant.  Moreover, the decay rates are maximal for momenta
parallel to the diagonals of the Brillouin zone and vanish along the
symmetry axes of the crystal, as illustrated 
in Fig.~\ref{fig:dampres}.
Our result (\ref{eq:dampres}) should be compared with
the Beliaev damping of sound in the ground state of the
superfluid Bose gas, which is proportional $k^3$  in two dimensions \cite{Sinner09}. In Eq.~(\ref{eq:dampres})
two extra powers of $k$ are generated 
by the strong momentum dependence of the squared scattering vertex
in Eq.~(\ref{eq:Wlong}).
This is similar to the XXZ ferromagnet \cite{Stephanovich2011},
where the hydrodynamic momentum dependence of the scattering vertex generates four extra powers of $ k $,
so that the damping behaves as $ k^7 $ at long wavelengths.
Note also that our result (\ref{eq:dampres}) for the damping at long wavelengths is not specific to the checkerboard model (\ref{eq:modelc}),
but generic to all models with the same symmetries,
since we expressed it in terms of the long-wavelength parameters $ c $ and $ m $.

%

Because the calculations presented
in this work are based on the bosonization of the spin operators
using the Holstein-Primakoff transformation \cite{Holstein40} and the expansion of the resulting
boson  Hamiltonian in powers of $1/S$, our results can only be expected to be quantitatively accurate for large spin $S$. An alternative bosonization of spin operators
proposed by Dyson \cite{Dyson56} and Maleev \cite{Maleev57} avoids the formal expansion of the Hamiltonian in powers of $1/S$, at the price of mapping the spin Hamiltonian onto a non-Hermitian boson Hamiltonian.
Although the Dyson-Maleev transformation has been successfully used to calculate magnon damping in quantum antiferromagnets \cite{Harris71,Kopietz90}, it  completely  fails in the present context because the non-Hermiticity of the Dyson-Maleev Hamiltonian  leads to negative
values of the product of scattering vertices in the diagrams shown in 
Fig.~\ref{fig:h4}~(b),
implying unphysical negative decay rates.
We are not aware of any other example for such a spectacular failure 
of the Dyson-Maleev transformation.

The anisotropy and chirality selectivity of
spontaneous magnon decay in altermagnets offers new possibilities of
identifying altermagnets experimentally. For example, since magnons propagating along the
symmetry axes of the crystal cannot spontaneously decay
while magnons propagating along the diagonals have the maximal possible decay rate,
the transport of magnetization or  heat along the symmetry axes is expected to be most efficient.
A popular method  for modeling  transport phenomena in magnets is based on the Landau-Lifshitz-Gilbert equation \cite{Gurevich96}, where the magnon damping is taken into account via  a phenomenological damping rate. Our result that spontaneous magnon decay in two-dimensional altermagnets
is anisotropic and chirality selective  cannot be straightforwardly
incorporated into the Gilbert damping phenomenology.

We thank the Deutsche Forschungsgemeinschaft (DFG, German Research Foundation) for financial support via TRR 288 - 422213477.

%
%


\begin{thebibliography}{99}
\bibitem{footnote_real_part}
Corrections from the real part of the magnon self-energy can be implicitly taken into account by redefining the long-wavelength parameters
$ c $ and $ 1 / m $ 
that characterize the magnon dispersion \eqref{eq:omegaexp}.
\end{thebibliography}
\end{document}